\documentclass[showpacs,superbib,10pt,twocolumn,aps]{revtex4}
\usepackage{graphicx}

\def\br{{\bf r}}
\def\bp{{\bf p}}
\def\by{{\bf y}}
\def\bq{{\bf q}}
\def\bz{{\bf z}}

\def\re{{\rm e}}
\def\bea{\begin{eqnarray}}
\def\eea{\end{eqnarray}}
\def\ben{\begin{equation}}
\def\een{\end{equation}}

\def\sss{\scriptscriptstyle\rm}

\def\ee{_{\sss ee}}

\def\2xc{_{\sss 2XC}}
\def\QC{^{\sss QC}}
\def\SC{^{\sss SC}}

\def\dulr{\underline{\underline \br}}
\def\dulp{\underline{\underline \bp}}
\def\dulq{\underline{\underline \bq}}
\def\dulz{\underline{\underline \bz}}

\begin{document}

\title{Semiclassical Correlation in Time-Dependent Density Matrix Functional Theory}
\author{A.K. Rajam, I. Raczkowska, N. T. Maitra}
\affiliation{Department of Physics and Astronomy, Hunter College and the Graduate Center of the City University of New York, 695 Park Ave, New York, NY 10065, USA}


\begin{abstract}
Lack of memory (locality in time) is a major limitation of almost all
present time-dependent density functional approximations. By using
semiclassical dynamics to compute correlation effects within a
density-matrix functional approach, we incorporate memory,
including initial-state dependence, as well as changing occupation numbers, and predict
 more observables in strong-field applications.
\end{abstract}

\maketitle 
The impact of time-dependent density functional theory
(TDDFT)~\cite{RG84,TDDFTbook} on calculations of excitation spectra
and response in atoms, molecules, and solids is evident in its
increasing use. In such applications a weak perturbation is applied to
the system beginning in its ground state, and usually the
exchange-correlation (xc) effects are treated with a ground-state
approximation. Generally the results are usefully accurate,
but specific cases (e.g. charge transfer excitations, optical response of
solids, etc) require improved approximations undergoing
intense research.

In principle, TDDFT also applies beyond the linear response regime,
but success has been slower.  There are three main reasons.  First,
many of the observables of interest are not simply related to the time-dependent one-body density: in addition to the approximation for the xc functional, new approximate ``observable functionals''
are needed  to extract the quantity of interest from the Kohn-Sham (KS) system. Even with 
an exact xc potential, they would remain elusive. These include
double-ionization probabilities and momentum-densities, and naive
approximations to these generally fail~\cite{WB06p,RGHM09}.
Second, lack of memory dependence in the usual xc approximations has been shown to
be often far more problematic than in the linear-response regime~\cite{MB01p, HMB02p}.  The exact functional depends on the history of the density as well as on the initial state.
Different initial states lead to fundamentally different xc potentials~\cite{MB01p}.
But no approximation today has
initial-state dependence, almost all neglect history-dependence, and all violate an exact condition on memory-dependence, derived in Ref.~\cite{MBW02}.
Third, a particularly severe difficulty is encountered when a system
starting in a wavefunction dominated by a single Slater determinant
(SSD), evolves to a state that fundamentally needs at least two SSDs
to describe it. This is the time-dependent (TD) analog of ground-state
static correlation, and arises in electronic quantum control
problems~\cite{MBW02,BWG05}, in ionization~\cite{RGHM09}, and in coupled electron-ion dynamics~\cite{LKQM06}. The TD KS system evolves the
occupied orbitals under a one-body Hamiltonian, remaining in an
SSD: the KS one-body density matrix is always idempotent (even with exact functionals), while, in contrast,
that of the true system develops eigenvalues (natural occupation
numbers) far from 0 or 1 in these applications.  The
exact xc potential  and observable functionals consequently develop complicated structure that is difficult to capture in approximations. 
 For
example, in Ref.~\cite{RGHM09}, a simple model of ionization in two-electron systems showed
that the momentum distribution computed directly from the exact KS
system contains spurious oscillations due to using a single, necessarily delocalized orbital, a non-classical 
 description of the essentially classical two-electron dynamics.
Ref.~\cite{MBW02} discussed the unusual and
non-intuitive xc potentials that arise in certain electronic
quantum-control problems, e.g. He 1s$^2 \to$ 1s2p.
If the overlap between the initial
and final states is targetted, the maximum that can be achieved is
0.5~\cite{BWG05}, while close to 0.98 is achieved for the true
interacting problem.

Recent pioneering strides in TD density-matrix functional theory (TDDMFT) show
this alternative approach can overcome some of the challenges of
adiabatic TDDFT in linear response~\cite{PGB07p}, e.g. adiabatic TDDMFT
functionals were shown to capture charge-transfer
excitations well. All one-body observables are directly obtained. However, adiabatic functionals bootstrapped from the usual ground-state DMFT 
 disappointingly cannot change occupation numbers unless some
unusual structural changes are made in the form of the
functional~\cite{PGB07p}. The first real-time TDDMFT calculations have been performed recently~\cite{RP09}, using an extra energy-minimizing procedure  at  each time that results in time-dependent occupation numbers.

In this Letter, we present a new approach to correlation in TDDMFT that makes a significant step in solving all the
problems mentioned above. 
We work within real-time TDDMFT 
and use a semiclassical
approximation for the correlation term in the equation of motion while
evaluating the other terms exactly quantum-mechanically. All one-body
observables are obtained directly, with correlation effects treated
semiclassically.  It is the first density-matrix(or density-)
functional approach that has initial-state dependence, with memory
naturally carried along by the classical trajectories, and the first
real-time approach that can change occupation numbers
significantly away from the adiabatic limit.  A heirarchy of semiclassical approximations for the
correlation term is discussed, decreasing in accuracy but also in
computational cost. On the first level, correlation is obtained
exactly to $O(\hbar)$, while at the lowest level quantum mechanics
enters only in the determination of the initial state, with the
dynamical correlation obtained via pure classical
evolution. Despite its simplicity, we demonstrate via a simple example that this latter approach yields time-dependent occupation numbers.

Semiclassics lies at the very foundations of the earliest density
functional theories that predate the rigorous DFT of
Refs.~\cite{TF2728p}. Its semiclassical origins were however
somewhat forgotten in the developments in the 1990's of
ground-state functionals, and only very recently have been
reawakened~\cite{ELCB08}. Until now, semiclassical
methods have not been applied to functional development in TDDFT nor in TDDMFT, although mean-field semiclassical methods have been used to approximate KS dynamics (e.g. Vlasov approaches to metal clusters in strong fields~\cite{DERS06p}). TDDMFT may equally be viewed as a ``phase-space-density functional theory'', as reflected for example in the relation between the one-body Wigner function $w(\br,\bp,t)$ and the spin-summed one-body density matrix $\rho_1(\br',\br,t)= N\sum_{\sigma_1..\sigma_N}\int d^3r_2..d^3r_N
\Psi^*(\br'\sigma_1,x_2..x_N,t)\Psi(\br\sigma_1,x_2..x_N,t)$:
\ben
w(\br,\bp,t) = \left(\frac{1}{2\pi}\right)^3\int d^3y\rho_1(\br-{\bf y}/2,\br + {\bf y}/2,t)e^{i\bp\cdot\by}
\label{eq:wigner}
\een 
(Here $x_i = (\br_i,\sigma_i)$ indicates spatial and spin coordinates, and atomic units are used throughout).
This observation suggests the utility of semiclassical approaches, as we shall see shortly.

All the one-body terms in the equation of motion for $\rho_1$ can be treated exactly, and for spin-unpolarized systems:
\begin{widetext}
\bea
\nonumber
i\dot{\rho_1}({\bf{r'}},{\bf{r}},t)&=&\left(-\nabla^2/2 + v({\bf{r}},t)+ \nabla'^2/2 - v({\bf{r'}},t)\right)\rho_1({\bf{r'}},{\bf{r}},t)\\
&+&\int d^3r_2 f\ee(\br,\br',\br_2)\left(n(\br_2,t)\rho_1(\br',\br,t)-\frac{1}{2}\rho_1(\br',\br_2,t)\rho_1(\br_2,\br,t)\right)
+\int d^3r_2 f\ee(\br,\br',\br_2)\rho_{2\sss C}(\br',\br_2,\br,\br_2,t)
\label{eq:rho1dot}
\eea
\end{widetext}
where we have decomposed the second-order density matrix, $\rho_2(\br',\br_2',\br,\br_2,t) = N(N-1)\sum_{\sigma_1..\sigma_N}\int d^3r_3..d^3r_N$
\newline
$\Psi^*(\br'\sigma_1,\br_2'\sigma_2,x_3..x_N,t)\Psi(\br\sigma_1,\br_2\sigma_2,x_3..x_N,t)
=
\rho_1(\br_1',\br_1,t)\rho_1(\br_2',\br_2,t)- \rho_1(\br_1',\br_2,t)\rho_1(\br_2',\br_1,t)/2 +\rho_{2\sss C}(\br_1',\br_2',\br_1,\br_2,t)$:
the first term is the non-interacting, uncorrelated product, the second term takes care of the Pauli principle at the uncorrelated level, 
and the third term is the correlation component, whose functional dependence on $\rho_1$ is unknown. 
In TDDMFT,  $\rho_2(\br',\br_2,\br,\br_2,t)$ is to be approximated
as a functional of $\rho_1$ and the initial interacting state $\Psi_0$, $\rho_2[\rho_1,\Psi_0]$. We have defined the electron-interaction kernel $f\ee(\br,\br',\br_2) = 1/\vert\br-\br_2\vert-  1/\vert\br'-\br_2\vert$.
If we evolve the $N$-electron interacting system in a
TD external potential $v(\br,t)$, that there is a one-to-one
mapping between $v(\br,t)$ and $\rho_1$ (or $w(\br,\bp,t)$) for a given
initial-state $\Psi_0(\br_1...\br_N)$ follows directly from the
Runge-Gross theorem. The mapping holds also for external vector
potentials~\cite{RGHM09} but we will focus on scalar potentials at
present. 
An immediate
advantage of replacing the coordinate-space density with the
density-matrix as basic variable is that it
directly gives the expectation value of {\it any} one-body operator:
no additional observable functionals are needed for
momentum-distributions or kinetic energies, for example. There is no
KS equivalent: because of idempotency of non-interacting density
matrices, it is impossible for a non-interacting system of electrons
to have the same phase-space density as a system of interacting
electrons.

Ideally, the approximation made for $\rho_2[\rho_1,\Psi_0]$ captures correlation, memory-dependence including initial-state dependence, and, most importantly for the quantum control and ionization applications mentioned earlier,  can yield time-dependent occupation numbers, $f_i(t)$, defined via the natural orbital decomposition $
\rho_1({\bf{r}},{\bf{r'}},t)=\sum_{i}f_i(t) \psi_i({\bf{r}},t) \psi^*_i({\bf{r'}},t)\;.
$

 The TDDMFT developments have so far been predominantly within linear response~\cite{PGB07p},
 investigating adiabatic functionals for $\rho_2$.
 Our approach computes the correlation component of $\rho_2$  via
 semiclassical dynamics, focussing on full dynamics (not linear response).
We propagate Eq.~(\ref{eq:rho1dot})
treating all terms except the last exactly quantum-mechanically. 
The last term is treated as a driving term: we approximate $\rho_{2\sss C} \approx\rho^{\sss SC}_{2\sss C}$, evaluated separately via  semiclassical dynamics, calculated from running classical trajectories in the $N$-body interacting phase-space.

Semiclassical methods construct an approximate quantum propagator
utilizing classical trajectory information alone. Although there are a variety of forms, the essential structure is a sum over classical trajectories:
\ben
\sum_{\rm cl. traj.} C_i(t)\re^{iS_i(t)/\hbar}
\een
where $S_i(t)$ is the classical action along the $i$th trajectory, and the prefactor $C_i(t)$ captures fluctuations around the classical path. 
Semiclassical approaches are able to capture quantum effects such as interference, zero-point energy, tunneling (to some extent), while generally scaling favorably with the number of degrees of freedom.  Based on classical
trajectories, intuition about the physical mechanisms
underlying the dynamics can be gained. Although mostly applied to  nuclear dynamics in molecules,
there have been  applications to electrons~\cite{Rostp}. 

Semiclassical formulae have been derived both from largely intuitive
arguments~(e.g. Ref.\cite{H81}), as well as from careful rigorous asymptotic analyses of the quantum propagator~(see e.g.~Refs.\cite{Schulmanp}) that satisfy TDSE to order $\hbar$.  
Miller~\cite{M74} showed the equivalence of different semiclassical representations within stationary-phase evaluation of the transformations.
The most popular is the Heller-Herman-Kluk-Kay~\cite{H81,KHD86, K05}, which is a ``semiclassical rigorization'' of Heller's frozen Gaussian approach, uniformly solving the TDSE to $O(\hbar)$. It is a sum over initial points in  ($N$-body) phase-space, $\dulz_0 \equiv (\dulr_0,\dulp_0) \equiv ((\br_1(0),\bp_1(0))...(\br_N(0),\bp_N(0)))$:
\begin{equation} 
\left(e^{-i\hat{H} t}\right)^{\rm SC} 
=\int \frac{d^{2M}\dulz_0}{(2\pi)^M} \vert \dulz_t \rangle
C_t(\dulz_0)e^{iS_t(\dulz_0)}\langle\dulz_0\vert
\label{eq:HK}
\end{equation} 
where:
$\dulz_t =(\dulr(t),\dulp(t))$ obeys Hamilton's equations
\ben
\dot{\dulr} =\dulp(t)\,,\;\;\;
\dot{\dulp}= -\nabla H(\dulr,\dulp,t)\;, 
\een
 $M=3N$ is the dimensionality of configuration space, and $S(\dulz_0,t)$ is
the classical action, $\int^t (T-V)dt$, for a trajectory which begins at the
phase space point $(\dulr_0,\dulp_0)$, reaching point $(\dulr_t,\dulp_t)$ at time
$t$. 
The state $\vert \dulz \rangle$ is a product of coherent states for each coordinate, labelled by their centers in phase-space: $\langle x\vert q,p\rangle = \left(\frac{\gamma}{\pi}\right)^{1/4} \exp(-\gamma(x-q)^2/(2) + ip(x-q))$, where $\gamma$ is a chosen width parameter.
The pre-exponential determinant $C_t(\dulz_0)$ accounts for fluctuations about the classical paths: when $\gamma$ is chosen identical for all particles,
$C_t(\dulz_0) =\left\vert
\frac{1}{2}\left(\frac{\partial
\br_t}{\partial\br_0}
+\frac{\partial\bp_t}{\partial\bp_0}
-i2\gamma\frac{\partial \br_t}{\partial\bp_0}
+\frac{i}{2\gamma}\frac{\partial
\bp_t}{\partial\br_0}\right)
\right\vert^{1/2}$.

Typically, the phase-space integral is performed via Monte-Carlo,
with initial conditions weighted by the initial wavepacket
$\langle\dulz_0\vert\Psi_0 \rangle$ .  Due to the evaluation of the prefactor $C$, the  numerical effort per trajectory scales as $N^3$; methods which neglect this scale more favorably as $N$ but at the cost of losing accuracy and semiclassical rigor. While Monte-Carlo integration scales as $\sqrt{N}$ for positive integrands, the phase-space integral can
be difficult to converge due to its oscillatory nature, especially for
many degrees of freedom and chaotic dynamics, and  so various sophisticated integral-filtering
techniques, or ``forward-backward'' (FB) methods~\cite{Makrip} have been formulated, allowing rigorous semiclassical calculations for up to 100 degrees of freedom~\cite{TW04}.

Applying Eq.~(\ref{eq:HK}) to propagate $\Psi$, then computing $\rho_2^{\rm SC}$ via integration constitutes our highest level of semiclassical heirarchy for correlation. We compute
\ben
\rho_1^{\sss SC}(\br',\br,t) = \frac{1}{N-1}\int\rho_2^{\sss SC}(\br',\br_2,\br,\br_2,t) d^3r_2
\label{eq:rho1sc}
\een
and extract
\bea
\nonumber
\rho_{2\sss C}^{\sss SC}(\br',\br_2,\br,\br_2,t) &=& \rho_{2}^{\sss SC}(\br',\br_2,\br,\br_2,t) - \rho_1^{\sss SC}(\br',\br,t)n^{\sss SC}(\br_2,t)\\
&+& \rho_1^{\sss SC}(\br',\br_2,t)\rho_1^{\sss SC}(\br_2,\br,t)/2
\label{eq:rho2csc}
\eea 
where  $n^{\sss SC}(\br_2,t)$ is the one-body density obtained from the semiclassical calculation, the diagonal of Eq.~(\ref{eq:rho1sc}).
Finally, this $\rho_{2\sss C}^{\sss SC}$ is input into
Eq.~(\ref{eq:rho1dot}) as a driving term. At this level, correlation
effects are exact to $O(\hbar)$, while all other effects are
quantum-mechanically exact. Difficulties with convergence of the
highly-oscillatory integral in Eq.~(\ref{eq:HK}), and the $N^3$ scaling
of the pre-factor, render this impractical for more than a few
electrons, to the point where, for many cases, little computational benefit is gained
over running the full quantum mechanics.

 Instead the FB idea lends itself particularly well to our purposes: we can take advantage of significant phase-cancellation between the propagation of $\Psi^*$ and that of $\Psi$ in calculating $\rho_2\SC$. 
Applying the semiclassical propagator Eq.~(\ref{eq:HK}) to both the $\Psi$ and $\Psi^*$ appearing in $\rho_2$, and doing some intermediate integrations via stationary phase, the second level in our semiclassical heirarchy is: 
\bea
\nonumber
\rho_2\SC(\br',\br_2,\br,\br_2)&=& \frac{N(N-1)}{(2\pi)^{3N+2}}\int d^{2M}\dulz_0d^3z'_{1,t}d^3z'_{2,t}e^{i(S(t) - S'(t))}\\
&\times&\mathcal{G}(\br',\br,\br_2;\bz'_{1,t}\bz'_{2,t}\bz_{1,t}\bz_{2,t})\langle\Psi_0\vert \dulz'_0\rangle\langle\dulz_0\vert\Psi_0\rangle
\label{eq:rho2sc}
\eea
where 
\newline
$\dulz_0' =
(\bq_{1,t}'(-t),\bp_{1,t}'(-t),\bq_{2,t}'(-t),\bp_{2,t}'(-t),\br_{3,0},\bp_{3,0}...\br_{N,0},\bp_{N,0})$ and 
$\mathcal{G}(\br',\br,\br_2; \bz'_{1,t}\bz'_{2,t}\bz_{1,t}\bz_{2,t}) = \langle\br'\vert\bz'_{1,t}\rangle^*\langle\br_2\vert\bz'_{2,t}\rangle^* \langle \br\vert \bz_{1,t}\rangle \langle \br_2\vert \bz_{2,t}\rangle$.
That is, an
initial phase-space point ${\dulq_0,\dulp_0}$ is classically evolved for
time $t$, when the phase-space points of the the first two particles
are shifted to $(\bq'_{t1},\bp'_{t1},\bq'_{t2},\bp'_{t2})$, before all
particles evolve back to time zero.  There is therefore significant
cancellation of phase ($S(t) - S'(t)$), that would generally result in
good convergence of Monte Carlo evaluation of this phase-space
integral, even for many electrons. Further, the product of the
numerically expensive pre-factors has been reasonably approximated to 1 for many
electrons. The true initial state appearing in Eq.~(\ref{eq:rho2sc})
is in practise approximated by a few (KS) SSD's, or by a high-level wavefunction
calculation, if a stationary state. Eqs.~(\ref{eq:rho1sc}) and~(\ref{eq:rho2csc}) are then
used to extract the semiclassical correlation component $\rho_{2\sss
  C}^{\sss SC}$, capturing interference and zero-point energy effects, that is then input into Eq.~(\ref{eq:rho1dot}) as a driving
term.

An even more simple prescription is obtained by neglecting the phase
and prefactor altogether: this yields the quasiclassical Wigner
method~\cite{H76p}, and can also be shown to result from a
linearization of the FB ~\cite{Makrip}:
\ben
w\QC_N({\underline{\underline \br}}, {\underline{\underline \bp}},t) =w_N\left({\underline{\underline \br}}_{-t}, {\underline{\underline \bp}}_{-t},t=0\right) 
\label{eq:QCW}
\een
from which, by integration, a quasiclassical approximation to the
correlation component of $\rho_2$ is obtained, and inserted as a
driving term into Eq.~\ref{eq:rho1dot}. This represents the lowest level of our semiclassical heirarchy:   in computing the correlation, while scaling classically with the system size, all interference is lost, quantum mechanics enters only in determining the initial Wigner function, and when the wavefunction
becomes delocalized, this approximation degrades. Nevertheless quasiclassical methods (even of the entire dynamics) have proven useful in analyzing electron ionization distributions~\cite{Agapi}. 

\begin{figure}[h]
\centering
\includegraphics[height=4.5cm,width=8.5cm]{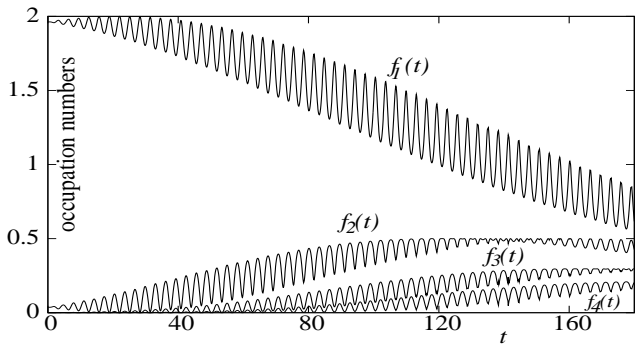}
\caption{Occupation numbers for the model system: quasiclassical correlation shown is exact, while the usual adiabatic TDDMFT approximations yield constant straight lines.}
\label{fig:evplot}
\end{figure}

Our prescription thus results in a semiclassical approximation for the
correlation component to $\rho_2$ in the equation of motion~Eq.~(\ref{eq:rho1dot}) for
$\rho_1$, all other terms of which are treated exactly
quantum-mechanically. 
But can our approach lead to time-dependent occupation numbers?
To illustrate this, we consider a simple model system, the two-electron Moshinsky atom~\cite{NSM08}:
\ben
\hat{H} = -\frac{1}{2}(\nabla_1^2 +\nabla_2^2) + \frac{k(t)}{2}(r_1^2 +r_2^2) + \lambda(r_1 - r_2)^2
\een
Although a poor model of a real atom, its purpose here is simply to
demonstrate that even the lowest level quasiclassical approximation to
correlation is able to capture changing occupation numbers. Its
harmonic nature makes it exactly solvable, and we apply a simple
sinusoidal force constant, $k(t) = 1 - 0.05\sin(2 t)$, that encourages
population transfer to the first accessible excited state (an
excitation in the center of mass coordinate), from the initial ground-state, a spin-singlet. Moreover, due to its
harmonic nature, the quasiclassical and semiclassical propagations are
exact.  Figure~\ref{fig:evplot} plots the occupation numbers, $f_i(t)$, of the spatial natural orbitals, obtained
from diagonalizing $\rho_1(\br,\br',t)$ at each time $t$. In striking
contrast, the results of the usual adiabatic approximations in TDDMFT
would yield constant occupation numbers. How well our approach works
for more realistic systems is currently being tested; this example
however illustrates that it certainly does not suffer from the
inability to change occupation numbers. Aside from the significance in quantum control problems, lack of time-dependent occupation numbers impacts observables, e.g. the momentum distributions are qualitatively incorrect~\cite{RGHM09}.

In summary, we have presented a semiclassical approach to correlation
in TDDMFT, that
(i) naturally captures history-dependence and initial-state dependence
(for the first time) at the semiclassical level, as memory is carried along with the classical trajectories composing $\rho_2\SC$, (ii) directly yields
all one-body observables, and (iii) changes
occupation numbers. Correlation is treated semiclassically, while all
other terms determining the density-matrix  are
exact. The highest semiclassical level yields correlation exactly to
$O(\hbar$), but will be impractical in many cases of interest; the approximate semiclassical treatment (Eq.~\ref{eq:rho2sc}) will still capture quantum many-body effects in a computationally efficient way. The simplest approximation (Eq.~\ref{eq:QCW}) scales classically, so is well worth investigating, especially since the other terms in the equation of motion for $\rho_1$ are treated exactly.
As there is no guarantee of $N$-representability of $\rho_1$ being preserved in the semiclassical dynamics (at least beyond $O(\hbar)$),
tests on more realistic systems than that presented here  are necessary.
Treating many of the challenging aspects of approximate
density-functional methods described earlier, it is a promising
approach to study electron dynamics in strong fields.

We thank Eric Suraud, Kieron Burke, and Peter Elliott for useful
conversations. Financial support from the National Science Foundation
grant CHE-0647913 and a Research Corporation Cottrell Scholar Award
(NTM) is gratefully acknowledged.

\end{document}